\documentclass[preprintnumbers,amsmath,amssymb,prd]{revtex4}

\usepackage{graphicx}
\usepackage{dcolumn}
\usepackage{bm}

\begin{document}

\preprint{DOI:10.1063/1.2162107}

\title{Covariant Irreducible Parametrization of Electromagnetic Fields \\in Arbitrary Spacetime}

\author{David Sundkvist}
 \altaffiliation[Now at ]{Swedish Institute of Space Physics, Uppsala, Sweden.}
 \email{davids@irfu.se}
\affiliation{
Laboratoire de Physique et Chimie de l'Environnement, CNRS, Orl\'eans, France}

\date{\today}

\begin{abstract}
We present a new unified covariant description of electromagnetic field properties
for an arbitrary space-time. We derive a complete set of irreducible components
describing a six-dimensional electromagnetic field from the Maxwell and metric
tensors using the symmetry group SL(2,C). For the special case of a flat space-time
metric the components are shown to correspond to the scalar invariants of the
electromagnetic field, the energy-momentum-stress tensor and in addition, three
new tensors expressing physical observables of rank two and four, respectively. We
make a physical interpretation of one of the new rank two tensors as describing a
classical intrinsic spin of the electromagnetic field. 
\end{abstract}

\maketitle

\section{Introduction}
The electromagnetic field in Maxwells theory is since long known to satisfy important symmetries and consequently satisfy conservation laws. 
For example the conservation law $T^{\alpha\beta}_{\phantom{\alpha\beta},\beta} =0$ where $T^{\alpha\beta}$ is the usual energy-momentum-stress tensor expresses the conservation of energy, momentum (Poynting) and stress densities of the electromagnetic field in a (local) Minkowski space. 
However, these are not the only symmetries and conserved quantities of physical relevance.
In Minkowski space the Maxwell equations are invariant 
under the important inhomogeneous Lorentz group, 
or as sometimes called, the Poincar\'e group. The flat spacetime of Minkowski space exhibits the maximum degree of symmetry one can obtain. 

Moving to the curved spacetime of general relativity many of the symmetries are lost, and conservation laws applies only locally. A group arising naturally in the study of general relativity is the group SL(2,C), the group of complex unimodular 2x2 matrices. 
There exist a homomorphism between SL(2,C) and the proper, orthochronous, homogeneous
Lorentz group L, 
\begin{equation}
\text{SL(2,C)}\rightarrow \text{L},
\end{equation}
reflecting the fact that SL(2,C) is the covering group of L and indicating how local Lorentz invariance is retained. 

In this paper we present a new unified treatment of the symmetries of the electromagnetic field in a general Riemannian spacetime by calculating the irreducible components of the covariant spectral density 
tensor. By performing a Fourier transform in time and considering the complex spectral densities we have further generalized to wave fields.

In section (\ref{seq:general}) we define the problem of finding the irreducible components by expressing the possible bilinear forms of the electromagnetic field in terms of the covariant spectral density. We then translate this to an equivalent problem under the group SL(2,C) by transforming to the tangent spinor space. After decomposing the spectral density spinor into irreducible components under SL(2,C) we transform back to the equivalent tensors in Riemann space. In section (\ref{seq:flat}) we study the important special case of a flat spacetime and calculate the components explicitly.
Some of them are shown to correspond to well known objects in Maxwells theory, while other components obtained have not previously been found in the litterature. Finally, in section (\ref{seq:conclusions}) we discuss our conclusions and present ideas for future work. 

\section{SL(2,C) theory of wave field spectral densities}\label{seq:general}

\subsection{Covariant spectral density tensor}
The fundamental physical object describing the electromagnetic field
is the electromagnetic field tensor $f_{\mu\nu}$ which is skew-symmetric
in its two indicies
\begin{equation}\label{eq:symmetry_of_fuv}
f_{\mu\nu}=-f_{\nu\mu}.
\end{equation}
It is comprised of the six components of the electromagnetic field.
We wish to categorize an electromagnetic
\emph{wave} field and therefore decompose the electric and magnetic fields in their spectral components.
Let $\bm f(t,\bm r)$ denote either the electric field $\bm E$ or magnetic field $\bm B$ at a
point in space and time
\begin{equation}
\bm f(t,\bm r)=\left(
\begin{array}{c}
f_x(t,\bm r)\\
f_y(t,\bm r)\\
f_z(t,\bm r)
\end{array}
\right).
\end{equation}
The Fourier transform in time, denoted by capital letters,
is then given by  
\begin{equation}\label{eq:fourier_transform}
\bm F(\omega,\bm r)=\int_{-\infty}^{\infty}\bm f(t,\bm r)e^{i\omega t}.
\end{equation}
In the following we assume that the fields $\bm E$, $\bm B$ and hence the tensor $F_{\mu\nu}$
are Fourier transformed according to (\ref{eq:fourier_transform}).
This implies that the the tensor $F_{\mu\nu}(\omega,\bm r)$ is now complex.
We form all possible bilinear forms of the electromagnetic field by constructing the outer product
\begin{equation}
S=F \otimes F^\dagger 
\end{equation}
which in tensor notation becomes
\begin{equation}\label{eq:covariant_spectral_density_tensor}
S_{\alpha\beta\gamma\delta}= F_{\alpha\beta}\overline{F_{\delta\gamma}}
\end{equation}
where the bar denotes complex conjugate. 
We call the complex tensor $S_{\alpha\beta\gamma\delta}$
in Eq.~(\ref{eq:covariant_spectral_density_tensor})
the \emph{covariant spectral density tensor}. 
From its definition 
and the skew-symmetry of $F_{\mu\nu}$, Eq.~(\ref{eq:symmetry_of_fuv}),
the symmetries of $S_{\alpha\beta\gamma\delta}$ follows as
\begin{subequations}
\label{eq:symmetries_of_S}
\begin{equation}\label{eq:antisymmetries_of_S}
S_{\alpha\beta\gamma\delta}=-S_{\beta\alpha\gamma\delta}=-S_{\alpha\beta\delta\gamma}=S_{\beta\alpha\delta\gamma}
\end{equation}
\begin{equation}\label{eq:conjugatesymmetries_of_S}
S_{\alpha\beta\gamma\delta}=\overline{S_{\gamma\delta\alpha\beta}}.
\end{equation}
\end{subequations}
Note that the decomposition in this section is not dependent on the particular transform used, but rather on the local structure and symmetry of the tensor $S_{\alpha\beta\gamma\delta}$.
This tensor have
$4^4=256$ complex components, but only 36 of these are independent due to the 
symmetries expressed in Eq.~(\ref{eq:antisymmetries_of_S}). This number is further 
decreased to half by the symmetry in Eq.~(\ref{eq:conjugatesymmetries_of_S}).
Accordingly, we find that $S_{\alpha\beta\gamma\delta}$ behaves as a six-dimensional hermitian matrix and
has 36 independent real valued components.
To find these  
we will find it convenient to reduce 
$S_{\alpha\beta\gamma\delta}$
into its irreducible parts. This will be done in the spinor formalism of SL(2,C).

\subsection{Spinor representation of spectral density}
Spinors arise in the representation theory of the group SL(2,C). For a short
review of the theory of spinors, see Appendix~\ref{app:spinors}, or a general reference such as \cite{carmeli1982a}.
According to Eq.~(\ref{eq:spinor_tensor_connection}) we find 
the spinor equivalent of the covariant spectral density
tensor 
to be
given by
\begin{eqnarray}\label{eq:S_spinor}
S_{AB^\prime CD^\prime EF^\prime GH^\prime}=\sigma^\alpha_{{AB^\prime}}\sigma^\beta_{{CD^\prime}}
	\sigma^\gamma_{{EF^\prime}}\sigma^\delta_{{GH^\prime}}S_{\alpha\beta\gamma\delta}
\end{eqnarray}
where the $\sigma^\mu_{AB'}$-matrices are the Infeld van der Waerden symbols, related to the
metric tensor $g_{\mu\nu}$ by Eq.~(\ref{eq:infeld_metric_connection}).
We call the spinor $S_{AB^\prime CD^\prime EF^\prime GH^\prime}$ the 
\emph{spectral density spinor}.
The symmetry condition Eq.~(\ref{eq:antisymmetries_of_S}) and Eq.~(\ref{eq:S_spinor}) implies 
\begin{subequations}\label{eq:S_spinor_symmetries_both}
\begin{equation}\label{eq:S_spinor_symmetries}
S_{AB^\prime CD^\prime EF^\prime GH^\prime}=-S_{CD^\prime AB^\prime EF^\prime GH^\prime}=-S_{AB^\prime CD^\prime GH^\prime EF^\prime}.
\end{equation}
Since $S_{\alpha\beta\gamma\delta}$ is complex, the spectral density spinor is not hermitian
in its indicies. 
To find how $S_{AB^\prime CD^\prime EF^\prime GH^\prime}$ transforms under complex conjugation
we consider Eq.~(\ref{eq:S_spinor}) and Eqs.~(\ref{eq:symmetries_of_S})
with the result
\begin{equation}\label{eq:S_spinor_complex_conjugate}
S_{AB^\prime CD^\prime EF^\prime GH^\prime}=\overline{S_{HG'FE'DC'BA'}}
=\overline S_{H'GF'ED'CB'A}.
\end{equation}
\end{subequations}

To find the decomposition of the spectral density spinor we make the observation that 
the equivalent spectral density tensor satisfies some of the same symmetries, 
Eq.~(\ref{eq:antisymmetries_of_S}),
as the Riemann tensor.
We therefore let \cite{witten1959a,carmeli1982a} inspire us
in the decomposition 
and 
make use of the identities Eqs.~(\ref{eq:useful_relation1},\ref{eq:useful_relation2}).
Writing the spectral density spinor in a symmetric form
\begin{eqnarray}\label{eq:gah}
&S_{AB' CD' EF' GH'}
=
\frac{1}{2}\left( 
S_{AB' CD' EF' GH'} -
S_{CB' AD' EF' GH'} 
\right)\nonumber\\
&+\frac{1}{2}\left(
S_{CB' AD' EF' GH'}-
S_{CD' AB' EF' GH'}
\right )\nonumber\\
&
\end{eqnarray}
and applying Eq.~(\ref{eq:useful_relation2}) gives
\begin{eqnarray}\label{eq:part_decomposition}
&S_{AB^\prime CD^\prime EF^\prime GH^\prime}= \frac{1}{2}
\left(
\varepsilon_{AC}~S_{IB^\prime\phantom{I}D^\prime EF^\prime GH^\prime}^{\phantom{IB\prime}I}+\right. \nonumber\\
&+\left.
S_{CJ'A\phantom{J'}EF'GH'}^{\phantom{CJ'A}J'}~\varepsilon_{B^\prime D^\prime}
\right)
\end{eqnarray}
Writing Eq.~(\ref{eq:part_decomposition}) in symmetric form and using 
Eq.~(\ref{eq:useful_relation2}) 
once more on each term 
results in
\begin{widetext}
\begin{eqnarray}\label{eq:S_Spinor_decomposition}
S_{AB^\prime CD^\prime EF^\prime GH^\prime}
=\frac{1}{4}\varepsilon_{AC}\left( 
S_{IB^\prime\phantom{I}D^\prime JF^\prime\phantom{J}H^\prime}^{\phantom{IB\prime}I\phantom{D^\prime JF^\prime}J}~\varepsilon_{EG} +
S_{IB^\prime\phantom{I}D^\prime EK^\prime G\phantom{K^\prime}}^{\phantom{IB\prime}I\phantom{D^\prime EK^\prime G}K^\prime}~\varepsilon_{F^\prime H^\prime} 
\right)\nonumber\\
+\frac{1}{4}\left(
S_{AI^\prime C\phantom{I^\prime}JF^\prime\phantom{J}H^\prime}^{\phantom{AI^\prime C}I^\prime\phantom{JF^\prime}J}~\varepsilon_{EG}+
S_{AI^\prime C\phantom{I^\prime}EK^\prime G}^{\phantom{AI^\prime C}I^\prime\phantom{EK^\prime G}K^\prime}~\varepsilon_{F^\prime H^\prime}
\right )
\varepsilon_{B^\prime D^\prime}
\end{eqnarray}
\end{widetext}
For the spinors in the last three terms in Eq.~(\ref{eq:S_Spinor_decomposition}) we introduce the notation
\begin{equation}\label{eq:gammaspinor}
\Gamma_{B'D'EG}=\frac{1}{4}S_{IB'\phantom{I}D'EK'G\phantom{K'}}^{\phantom{IB'}I\phantom{D'EK'G}K'}
\end{equation}
\begin{equation}\label{eq:deltaspinor}
\Delta_{ACF^\prime H^\prime}=\frac{1}{4}S_{AI^\prime C\phantom{I^\prime}JF^\prime\phantom{J}H^\prime}^{\phantom{AI^\prime C}I^\prime\phantom{JF^\prime}J}
\end{equation}
\begin{equation}\label{eq:sigmaspinor}
\Sigma_{ACEG}=\frac{1}{4}S_{AI^\prime C\phantom{I^\prime}EK^\prime G}^{\phantom{AI^\prime C}I^\prime\phantom{EK^\prime G}K^\prime}.
\end{equation}
From Eq.~(\ref{eq:S_spinor_symmetries}) and Eqs.~(\ref{eq:gammaspinor}-\ref{eq:sigmaspinor}) 
it follows that the symmetries of the decomposed spinors are
\begin{equation}\label{eq:Gammasymmetry}
\Gamma_{B'D'EG}=\Gamma_{D'B'EG}=\Gamma_{B'D'GE}
\end{equation}
\begin{equation}\label{eq:Deltasymmetry}
\Delta_{ACF'H'}=\Delta_{CAF'H'}=\Delta_{ACH'F'}
\end{equation}
\begin{equation}\label{eq:Sigmasymmetry}
\Sigma_{ACEG}=\Sigma_{CAEG}=\Sigma_{ACGE}.
\end{equation}
For the first term in Eq.~(\ref{eq:S_Spinor_decomposition}) we obtain
\begin{equation}\label{eq:sigmaconjugate}
\frac{1}{4}S_{IB'\phantom{I}D'JF'\phantom{J}H'}^{\phantom{IB'}I\phantom{D'JF'}J}=
\frac{1}{4}\overline{S_{FJ'H\phantom{J'}BI'D}^{\phantom{FJ'H}J'\phantom{BI'D}I'}}=
\overline\Sigma_{F'H'B'D'}
\end{equation}
where we have used Eqs.~(\ref{eq:S_spinor_symmetries_both}) and Eq.~(\ref{eq:sigmaspinor}).
The $\Gamma$ and $\Delta$-spinors in the second and third terms of Eq.~(\ref{eq:S_Spinor_decomposition})
both contain mixed indicies 
and do not satisfy a relation similar to Eq.~(\ref{eq:sigmaconjugate}). Instead they transform under 
complex conjugation as
\begin{equation}\label{eq:gammaconjugate}
\Gamma_{B'D'EG}=\overline{\Gamma_{E'G'BD}}=\overline\Gamma_{EGB'D'}
\end{equation}
\begin{equation}\label{eq:deltaconjugate}
\Delta_{ACF'H'}=\overline{\Delta_{FHA'C'}}=\overline\Delta_{F'H'AC}
\end{equation}
where we have used Eqs.~(\ref{eq:S_spinor_symmetries_both}).
Hence, Eq.~(\ref{eq:S_Spinor_decomposition}) can be written 
{\setlength\arraycolsep{2pts}
\begin{eqnarray}
\label{eq:decomposed_S_spinor}
S_{AB^\prime CD^\prime EF^\prime GH^\prime}
&=&\nonumber\\
\varepsilon_{AC}(
\overline\Sigma_{F'H'B'D'}~\varepsilon_{EG} + 
\Gamma_{B'D'EG}~\varepsilon_{F'H'}
)
+&&\nonumber\\
(
\Delta_{ACF'H'}~\varepsilon_{EG}+
\Sigma_{ACEG}~\varepsilon_{F^\prime H^\prime}
)
\varepsilon_{B^\prime D^\prime}.
\end{eqnarray}
}

\subsection{Number of independent parameters}
\label{seq:num_indep_para}
From 
Eq.~(\ref{eq:decomposed_S_spinor}) we may find the number of
independent parameters needed to fully describe an electromagnetic 
wave field. Introducing the notation 
$(1,2,3)=(00,01=10,11)$
we can write the components of the $\Sigma_{ACEG}$-spinor as
$\Sigma_{0000}=\Sigma_{11}$, $\Sigma_{1000}=\Sigma_{0100}=\Sigma_{21}$,
$\Sigma_{1100}=\Sigma_{31}$,
and similarily for the other components. 
From the symmetries in Eq.~(\ref{eq:Sigmasymmetry}) 
we can then view the components of 
$\Sigma_{ACEG}$ as a matrix $\Sigma$ 
with nine independent complex components
\begin{equation}
\bm\Sigma=
\left( 
\begin{array}{ccc}
\Sigma_{11} & \Sigma_{12} & \Sigma_{13} \\
\Sigma_{21} & \Sigma_{22} & \Sigma_{23} \\
\Sigma_{31} & \Sigma_{32} & \Sigma_{33} \\
\end{array}
\right)
\end{equation}
or equivalently 18 real independent components.
The $\Delta_{ACF'H'}$ and $\Gamma_{B'D'EG}$ 
spinors in addition to Eq.~(\ref{eq:Gammasymmetry}) 
and Eq.~(\ref{eq:Deltasymmetry}) also satisfies
Eq.~(\ref{eq:deltaconjugate}) and Eq.~(\ref{eq:gammaconjugate}).
Hence the corresponding matricies behaves like hermitian $3 \times 3$ matrices
$\Delta_{ij}=\overline\Delta_{ji}$, $\Gamma_{ij}=\overline\Gamma_{ji}$ where $(i,j)\in(1,2,3)$,
i.e $\Delta=\Delta^\dagger$, $\Gamma=\Gamma^\dagger$. 
Therefore the spinor $\Delta_{ACF'H'}$ has 
three real components $\Delta_{11}, \Delta_{22}, \Delta_{33}$ and three complex
components $\Delta_{12}, \Delta_{13}, \Delta_{23}$, in total nine real
independent components. The same applies to the $\Gamma_{B'D'EG}$ spinor.
In total we find $18+9+9=36$ real independent components describing the 
wave electromagnetic field in the spinor formulation, in accordance with the 
discussion following Eq.~(\ref{eq:symmetries_of_S}).

\subsection{Irreducible spinor representation}
Eq.~(\ref{eq:decomposed_S_spinor}) is still not on an irreducible form. To find the first 
irreducible component we form the contracted spinor
\begin{eqnarray}
&S_{AB'CD'}=S^{EF'}_{\phantom{EF'}AB'EF'CD'}=\nonumber\\
&\varepsilon_{AC}\varepsilon_{B'D'}\frac{1}{2}(\lambda+\lambda^*)
-\Gamma_{B'D'AC} - \Delta_{ACB'D'}
\end{eqnarray}
where we have 
defined $\lambda$ as the trace of the $\Sigma$ spinor
\begin{equation}\label{eq:lambdaspinorscalar}
\lambda=\Sigma_{AE}^{\phantom{AE}AE}
\end{equation}
and used 
\begin{equation}\label{eq:lambdarelation1}
\Sigma_{AE\phantom{E}C}^{\phantom{AE}E}=\frac{\lambda}{2}\varepsilon_{AC} 
\end{equation}
\begin{equation}
\overline\Sigma_{D'F'\phantom{F'}B'}^{\phantom{F'D'}F'}=\frac{\lambda^*}{2}\varepsilon_{D'B'}. 
\end{equation}

The trace $S$ of the contracted spinor $S_{AB'CD'}$
\begin{eqnarray}
&S=S_{AB'}^{\phantom{AB'}AB'}=\varepsilon^{AG}\varepsilon^{B'H'}S_{AB'GH'}=\nonumber\\
&2(\lambda+\lambda^*)
-\Gamma_{B'\phantom{B'}A}^{\phantom{B'}B'\phantom{A}A} - \Delta_{A\phantom{A}B'}^{\phantom{A}A\phantom{B'}B'}=
2(\lambda+\lambda^*)
\end{eqnarray}
is thus real. We now form the spinor 
\begin{equation}
M_{AB'CD'}=S_{AB'CD'}-\frac{1}{4}g_{AB'CD'}S
\end{equation}
which is traceless by construction and hence irreducible. 
From its definition we find the relations
\begin{equation}\label{eq:dspinor}
M_{AB'CD'}=-\Gamma_{B'D'AC} - \Delta_{ACB'D'}
\end{equation}
\begin{equation}
M_{AB'CD'}=M_{CD'AB'}
\end{equation}
\begin{equation}
M_{AB'CD'}=\overline{M_{BA'DC'}}.
\end{equation}
Calculating the trace of the terms involving the $\Gamma$ and $\Delta$ spinors in Eq.~(\ref{eq:decomposed_S_spinor})
we find
\begin{eqnarray}
M_{AB'CD'}=
\varepsilon^{EG}\varepsilon^{F'H'}(\varepsilon_{GA}\varepsilon_{F'D'}\Gamma_{H'B'EC}\nonumber\\
+\varepsilon_{EC}\varepsilon_{H'B'}\Delta_{GAF'D'}).&
\end{eqnarray}
We therefore write these terms in a form symmetric and anti-symmetric in exchange of the first and second against 
the third and forth pair of indicies with the result
\begin{widetext}
\begin{eqnarray}\label{eq:gammadeltadecompos}
\varepsilon_{AC}\varepsilon_{F'H'}\Gamma_{B'D'EG} +
\varepsilon_{EG}\varepsilon_{B'D'}\Delta_{ACF'H'}=\nonumber\\
\frac{1}{2}(
\varepsilon_{EG}\varepsilon_{B'D'}D_{AF'CH'}-\varepsilon_{AC}\varepsilon_{F'H'}D_{EB'GD'}
-\varepsilon_{AC}\varepsilon_{F'H'}M_{EB'GD'}-\varepsilon_{EG}\varepsilon_{B'D'}M_{AF'CH'}
)
\end{eqnarray}
\end{widetext}
where we have also defined 
\begin{equation}\label{eq:espinor}
D_{AB'CD'}=-\Gamma_{B'D'AC} + \Delta_{ACB'D'}
\end{equation}
which has the same symmetries as $M_{AB'CD'}$ and is also traceless.

It remains to find the irreducible parts of the $\Sigma$ spinors in Eq.~(\ref{eq:decomposed_S_spinor}).
To this end we write $\Sigma$ in a form utilizing the symmetries expressed in Eq.~(\ref{eq:Sigmasymmetry})
\begin{widetext}
\begin{eqnarray}\label{eq:Sigmaexpansion}
\Sigma_{ABCD}&=&\Psi_{ABCD}
+\frac{1}{3!}(\Sigma_{ABCD}-\Sigma_{DABC})
+\frac{1}{3!}(\Sigma_{ABCD}-\Sigma_{CDAB})
+\frac{1}{3!}(\Sigma_{ABCD}-\Sigma_{BCDA})+\nonumber\\
&&+\frac{1}{3!}(\Sigma_{ABCD}-\Sigma_{DBAC})
+\frac{1}{3!}(\Sigma_{ABCD}-\Sigma_{ACDB})
\end{eqnarray}
\end{widetext}
where 
\begin{eqnarray}\label{eq:psispinor}
\Psi_{ABCD}&=&\frac{1}{3!}(\Sigma_{ABCD}+\Sigma_{DABC}+\Sigma_{CDAB}+\nonumber\\
&&+\Sigma_{BCDA}+\Sigma_{DBAC}+\Sigma_{ACDB})
\end{eqnarray}
is a completely symmetric spinor, 
\begin{equation}
\Psi_{ABCD}=\Psi_{BACD}=\Psi_{ABDC}=\Psi_{ACBD}.
\end{equation}

Using Eqs.~(\ref{eq:useful_relation2}) and (\ref{eq:lambdarelation1}) 
on the remaining terms
of Eq.~(\ref{eq:Sigmaexpansion}) we find
\begin{equation}\label{eq:Sigmadecomposition}
\Sigma_{ABCD}=\Psi_{ABCD}+\frac{\lambda}{6}(\varepsilon_{AC} \varepsilon_{BD} + \varepsilon_{AD} \varepsilon_{BC})+\xi_{ABCD}
\end{equation}
where we have defined the spinor
\begin{equation}\label{eq:xispinor}
\xi_{ABCD}=\frac{1}{6}(\Sigma_{ABCD}-\Sigma_{CDAB})
\end{equation}
satisfying the symmetries
\begin{equation}
\xi_{ABCD}=\xi_{BACD}=\xi_{ABDC}=-\xi_{CDAB}.
\end{equation}
Inserting Eq.~(\ref{eq:Sigmadecomposition}) and its complex conjugate equivalent into 
(\ref{eq:decomposed_S_spinor}) 
and using (\ref{eq:gammadeltadecompos})
finally gives 
\begin{widetext}
\begin{eqnarray}\label{eq:irreduciblespinorrepres}
S_{AB'CD'EF'GH'}&=&C_{AB'CD'EF'GH'}+\nonumber\\
&&+\frac{\lambda}{6}\varepsilon_{B'D'}\varepsilon_{F'H'}(\varepsilon_{AE}\varepsilon_{CG}+\varepsilon_{AG}\varepsilon_{CE})+
\frac{\lambda^*}{6}\varepsilon_{AC}\varepsilon_{EG}(\varepsilon_{F'B'}\varepsilon_{H'D'}+\varepsilon_{F'D'}\varepsilon_{H'B'})+\nonumber\\
&&+\frac{1}{2}(
\varepsilon_{EG}\varepsilon_{B'D'}D_{AF'CH'}-\varepsilon_{AC}\varepsilon_{F'H'}D_{EB'GD'}
-\varepsilon_{AC}\varepsilon_{F'H'}M_{EB'GD'}-\varepsilon_{EG}\varepsilon_{B'D'}M_{AF'CH'}
)
\end{eqnarray}
\end{widetext}
The spinor $C_{AB'CD'EF'GH'}$ is defined by 
\begin{widetext}
\begin{equation}\label{eq:Cspinor}
C_{AB'CD'EF'GH'}=A_{AB'CD'EF'GH'}+B_{AB'CD'EF'GH'}
\end{equation}
where
\begin{equation}\label{eq:Aspinor}
A_{AB'CD'EF'GH'} = \Psi_{ACEG}\varepsilon_{F'H'}\varepsilon_{B'D'}+ \overline \Psi_{F'H'B'D'}\varepsilon_{AC}\varepsilon_{EG}
\end{equation}
and
\begin{equation}\label{eq:Bspinor}
B_{AB'CD'EF'GH'} = \xi_{ACEG}\varepsilon_{F'H'}\varepsilon_{B'D'} + \overline \xi_{F'H'B'D'}\varepsilon_{AC}\varepsilon_{EG}
\end{equation}
\end{widetext}

The decomposition of $C_{AB'CD'EF'GH'}$ into A and B is equivalent to a decomposition of the spinor into real and imaginary parts, $A_{AB'CD'EF'GH'}=\Re (C_{AB'CD'EF'GH'})$ and $B_{AB'CD'EF'GH'}=\Im (C_{AB'CD'EF'GH'})$.
The spinor $A_{AB'CD'EF'GH'}$ is identically zero when contracted over the first and third pair of indicies
\begin{equation}
A_{\phantom{EF'}AB'EF'CD'}^{EF'}=0
\end{equation}
due to its symmetries and real valuedness
which is readily verified after some straightforward algebra.
The spinor $B_{AB'CD'EF'GH'}$ can be contracted to form
\begin{equation}\label{eq:Uspinor}
\Upsilon_{AB'CD'} = B_{\phantom{EF'}AB'EF'CD'}^{EF'}
\end{equation}
satisfying
\begin{equation}
\Upsilon_{AB'CD'} = -\Upsilon_{CD'AB'} =\overline \Upsilon_{CD'AB'}
\end{equation}
and which is traceless in turn
\begin{equation}
\Upsilon_{\phantom{AB'CD'}AB'CD'}^{AB'CD'}=0
\end{equation}

Equations (\ref{eq:irreduciblespinorrepres}-\ref{eq:Uspinor}) is comprised only of scalars or traceless spinors and
hence is the sought irreducible representation of the spectral density spinor. In summary the decomposition is 
\begin{eqnarray}\label{eq:spinordecomptotal}
S_{AB'CD'EF'GH'} =  A_{AB'CD'EF'GH'}~\oplus~\Upsilon_{AB'CD'}\nonumber\\
\oplus~M_{AB'CD'}~\oplus D_{AB'CD'}~\oplus~\lambda.
\end{eqnarray}
Counting components we find that Eq.~(\ref{eq:Cspinor}) has 5 plus 3 complex components 
from $\Psi$ and $\xi$ respectively,
which follows from the symmetries. From the discussion in section~(\ref{seq:num_indep_para}) we know
that the $\Gamma$ and $\Delta$ spinors has 9 real components each which implies that
M and D have 18 real components in total. 
Together with the complex invariant $\lambda$ it adds up to 36 independent real components, in agreement with
the above discussion.

\subsection{Relation to the Riemann tensor}

We may note that if the covariant spectral density tensor was real and satisfied
$S_{\alpha\beta\gamma\delta}=S_{\gamma\delta\alpha\beta}$
instead of Eq.~(\ref{eq:conjugatesymmetries_of_S}) it would 
satisfy the same symmetries
as the Riemann tensor, $S_{\alpha\beta\gamma\delta}=R_{\alpha\beta\gamma\delta}$.
Indeed in such a case $D_{AB'CD'}=0$, $\xi_{AB'CD'}=0$, $\Gamma_{AB'CD'}=\overline \Delta_{CD'AB'}$ 
so that 
$M_{AB'CD'}=2\Delta_{AB'CD'}$
would be the traceless Ricci spinor,
$A_{AB'CD'EF'GH'}=\Psi_{ACEG}\varepsilon_{F'H'}\varepsilon_{B'D'}+ \overline \Psi_{F'H'B'D'}\varepsilon_{AC}\varepsilon_{EG}
$
the traceless Weyl spinor and 
$\lambda=\lambda^*=R$ would be the Ricci scalar curvature.
Hence Eq.~(\ref{eq:irreduciblespinorrepres}) would be completely analogous to the irreducible spinor representation of the Riemann curvature tensor.

\subsection{Irreducible tensor representation}
It is interesting to transform Eq.~(\ref{eq:irreduciblespinorrepres}) to
its tensor form, using the relation between a spinor and its equivalent tensor, given
by Eq.~(\ref{eq:tensor_spinor_connection}). In summary, we can write this decomposition of the covariant spectral density tensor into its irreducible components
symbolically as
\begin{equation}\label{eq:irrtensorsymb}
S_{\alpha\beta\gamma\delta}=A_{\alpha\beta\gamma\delta} \oplus \Upsilon_{\alpha\beta} \oplus M_{\alpha\beta} \oplus D_{\alpha\beta} \oplus \lambda,
\end{equation}
with the number of independent components 10+6+9+9+2=36.
In connection to this it is worthwhile noting that the complex scalar invariant $\lambda$, 
the only quantity of the electromagnetic field different observers agree on,
satisfies
\begin{equation}
\Re~(\lambda)=\frac{1}{4}F_{\mu\nu}F^{\mu\nu}
\end{equation}
\begin{equation}
\Im~(\lambda)=-\frac{1}{8}F_{\mu\nu}{}^*F^{\mu\nu}
\end{equation}
where ${}^*F^{\mu\nu}$ is the dual tensor
\begin{equation}
{}^*F^{\mu\nu}=\frac{1}{2}\varepsilon^{\mu\nu\rho\sigma}F_{\rho\sigma}.
\end{equation}
Hence, we recover the two scalar invariants of electromagnetic field theory as the
real and imaginary parts of the complex scalar invariant of the covariant spectral density
tensor.

\section{Special case: Flat spacetime}\label{seq:flat}
In this section we treat the important special case of a flat spacetime
in the absence of gravitation, with the prescribed metric given by 
\begin{equation}\label{eq:minkowskimetric}
g_{\mu\nu}=\left(
\begin{array}{rrrr}
1 & 0 & 0 & 0\\
0 & -1 & 0 & 0\\
0 & 0 & -1 & 0\\
0 & 0 & 0 & -1
\end{array}
\right).
\end{equation}
The electromagnetic field tensor is representated by
\begin{equation}
F_{\mu\nu}=\left(
\begin{array}{rrrr}
0 & -E_x & -E_y & -E_z\\
E_x & 0 & B_z & -B_y\\
E_y & -B_z & 0 & B_x\\
E_z & B_y & -B_x & 0
\end{array}
\right)
\end{equation}
in natural units where we have put $c=1$.

The solution to Eq.~(\ref{eq:infeld_metric_connection}) for the given 
Minkowskian metric Eq.~(\ref{eq:minkowskimetric})
gives the Infeld van der Waerden symbols
\begin{equation}\label{eq:waerdenflat}
\begin{array}{cc}
\sigma^0_{AB'}=\frac{1}{\sqrt 2}
\left( 
\begin{array}{cc}
1 & 0  \\
0 & 1  \\
\end{array}
\right)
&
\sigma^1_{AB'}=\frac{1}{\sqrt 2}
\left( 
\begin{array}{cc}
0 & 1  \\
1 & 0  \\
\end{array}
\right)
\\ \\
\sigma^2_{AB'}=\frac{1}{\sqrt 2}
\left( 
\begin{array}{cc}
0 & i  \\
-i & 0  \\
\end{array}
\right)
&
\sigma^3_{AB'}=\frac{1}{\sqrt 2}
\left( 
\begin{array}{cc}
1 & 0  \\
0 & -1  \\
\end{array}
\right)
\end{array}
\end{equation}
proportional to the Pauli matrices \cite{carmeli1982a}.
We now follow the following scheme: 
\begin{enumerate}
\item 
Form the spectral density tensor Eq.~(\ref{eq:covariant_spectral_density_tensor}).
\item 
Calculate the spectral density spinor Eq.~(\ref{eq:S_spinor}) using Eq.~(\ref{eq:waerdenflat}).
\item 
Calculate $\Gamma, \Sigma, \Delta$ using Eq.~(\ref{eq:gammaspinor}-\ref{eq:sigmaspinor}). 
\item 
Calculate the irreducible spinor components $\lambda$ from Eq.~(\ref{eq:lambdaspinorscalar}), $M_{AB'CD'}$ from Eq.~(\ref{eq:dspinor}), $D_{AB'CD'}$ from Eq.~(\ref{eq:espinor}), $\Psi$ from Eq.~(\ref{eq:psispinor}) and $\xi$ using Eq.~(\ref{eq:xispinor}).
\item 
Transform the spinors obtained to their irreducible tensor counterparts using Eq.~(\ref{eq:tensor_spinor_connection}) and the spinorally contravariant form of Eq.~(\ref{eq:waerdenflat}).
\end{enumerate}

\subsection{Fundamental objects as irreducible components}
By following the above scheme we find the SL(2,C) 
irreducible tensor components that comprise the spectral density tensor according to Eq.~(\ref{eq:irrtensorsymb}). The first tensor, the complex scalar invariant $\lambda$ is found to be 
\begin{equation}\label{eq:lambdascalarflat}
\lambda = \frac{1}{2}
\left[  
	|{\bf E}|^2-|{\bf B}|^2
\right] + i\Re
		\left(  
			{\bf E}\cdot{\bf B^*}
		\right)
\end{equation}
We identify the real and imaginary part as the scalar Lagrangian invariant and the pseudo scalar invariant (here expressed in complex form since the fields are Fourier transformed). As is well known, these are the only two invariants that exist in Maxwells theory. This is verified here by the fact that only scalars are true invariants under a general spin frame transformation. 

Calculating the components of the first rank two tensor $M_{\alpha\beta}$ we find
\begin{equation}\label{eq:Dtensor}
M_{\alpha\beta}=\left(
\begin{array}{rrrr}
\sigma & P_x & P_y & P_z\\
P_x & T_{xx} & T_{xy} & T_{xz}\\
P_y & T_{yx} & T_{yy} & T_{yz}\\
P_z & T_{zx} & T_{zy} & T_{zz}
\end{array}
\right)
\end{equation}
where we have used the notation
\begin{equation}\label{eq:energydensity}
\sigma=\frac{1}{2}\left(|{\bf E}|^2+|{\bf B}|^2\right)
\end{equation}
\begin{equation}
P_x=\frac{1}{2}\left[\Re\left(E_yB_z^*\right)-\Re\left(B_y^*E_z\right)\right]
\end{equation}
\begin{equation}
P_y=\frac{1}{2}\left[\Re\left(E_zB_x^*\right)-\Re\left(B_z^*E_x\right)\right]
\end{equation}
\begin{equation}
P_z=\frac{1}{2}\left[\Re\left(E_xB_y^*\right)-\Re\left(B_x^*E_y\right)\right]
\end{equation}
\begin{equation}
T_{ij}=-\Re \left[E_iE_j^*+B_iB_j^*\right] + \delta_{ij}\sigma.
\end{equation}
We immediately recognize Eq.~(\ref{eq:Dtensor}) as the energy-momentum-stress tensor in Maxwells theory, here obtained as one of the irreducible components of the spectral density tensor under SL(2,C). We identify as usual $\sigma$ as the energy density of the electromagnetic field, $(P_x,P_y,P_z)=\frac{1}{2}\Re(\bf E \times \bf B^*)$ as the complex Poynting vector and $T_{ij}$ as the three-dimensional Maxwell stress tensor. 

Turning our attention to the second rank two irreducible tensor $D_{\alpha\beta}$ we find analogously 
\begin{equation}\label{eq:Etensor}
D_{\alpha\beta}=\left(
\begin{array}{rrrr}
K & Q_x & Q_y & Q_z\\
Q_x & U_{xx} & U_{xy} & U_{xz}\\
Q_y & U_{yx} & U_{yy} & U_{yz}\\
Q_z & U_{zx} & U_{zy} & U_{zz}
\end{array}
\right)
\end{equation}
where we have used the notation
\begin{equation}\label{eq:spindensity}
K=-\Im(\bf E \cdot \bf B^*)
\end{equation}
\begin{equation}\label{eq:spindensityvector}
{\bf Q}=-\frac{i}{2} \left(\bf E \times \bf E^* + \bf B \times \bf B^*\right)
\end{equation}
\begin{equation}
U_{ij}=-\Im \left[E_i^*B_j - B_i^*E_j\right] + \delta_{ij} K.
\end{equation}
We identify the $D_{00}$ component as the imaginary part of the pseudoscalar invariant of the electromagnetic field. Since also e.g $Q_x=-\Im (E_zE_y^* + B_zB_y^*)$ we note that all components of the tensor $D_{\alpha\beta}$ can be written as imaginary parts. In analogy with $\sigma$, the energy density Eq.~(\ref{eq:energydensity}), we may denote Eq.~(\ref{eq:spindensity}) the spin density of the electromagnetic field. This is because ${\bf Q}$, Eq.~(\ref{eq:spindensityvector}), is only different from zero in the case when the electromagnetic field is elliptically (i.e not linearly) polarized. We call ${\bf Q}$ the ``spin flux density''.

Finally we calculate the components of the remaining rank four tensor $C_{\alpha\beta\gamma\delta}$.
Its real part $A_{\alpha\beta\gamma\delta}$ is traceless
\begin{equation}
A^{\gamma}_{\phantom{\gamma}\alpha\gamma\beta} = 0
\end{equation}
when contracted. 
This tensor has ten independent components. In order to write this rank four tensor on a compact form, we introduce the bijective mapping of the index pairs (1,0)$\leftrightarrow$(1), (2,0)$\leftrightarrow$(2), (3,0)$\leftrightarrow$(3), (3,2)$\leftrightarrow$(4), -(3,1)$\leftrightarrow$(5), (2,1)$\leftrightarrow$(6). We can then construct the column sixtors $A_{k}$, where for example the first vector is $A_{1}=(A_{1010},A_{1020},A_{1030},A_{1032},-A_{1031},A_{1021})^T$. The tensor $A_{\alpha\beta\gamma\delta}$ is then equivalent to the 6x6 matrix $\textrm{A}_{kl}=(A_1~A_2~A_3~A_4~A_5~A_6)$. In index notation it becomes
\begin{equation}\label{eq:Atensor}
A_{\alpha\beta\gamma\delta}\leftrightarrow\left(
\begin{array}{rr}
E & F\\
F & -E
\end{array}
\right)
\end{equation}
where the two block matrices
\begin{equation}
E=E_{ij}=-\frac{1}{2}\Re(E_i^*E_j-B_i^*B_j)+\frac{2}{6}\Re(\lambda)\delta_{ij}
\end{equation}
and
\begin{equation}
F=F_{ij}=-\frac{1}{2}\Re(E_i^*B_j+B_i^*E_j)+\frac{2}{6}\Im(\lambda)\delta_{ij}
\end{equation}
each contains five independent components.

The contracted imaginary part of $C_{\alpha\beta\gamma\delta}$ behaves as an anti-symmetric hermitian matrix and have the components
\begin{equation}\label{eq:upsilontensor}
\Upsilon_{\alpha\beta}=\left(
\begin{array}{rrrr}
0 & \cdot & \cdot & \cdot\\
S_x & 0 & \cdot & \cdot\\
S_y & V_{yx} & 0 & \cdot\\
S_z & V_{zx} & V_{zy} & 0
\end{array}
\right)
\end{equation}
where
\begin{equation}
S=(S_x,S_y,S_z)=-\frac{i}{3}\Im({\bf E}\times {\bf B}^*)
\end{equation}
is the imaginary part of the complex Poynting vector and 
\begin{equation}
V_{ij}=-\frac{i}{3}\Im(E_i^*E_j-B_i^*B_j)
\end{equation}
is a three-tensor. 

\subsection{Physical interpretation}

Interestingly, as we noted above, the real part of the scalar invariant Eq.~(\ref{eq:lambdascalarflat}) of the electromagnetic field would play the role of the Ricci scalar curvature in the case of real valued fields. Similarily the energy-momentum-stress tensor $M_{\alpha\beta}$ would palsolay the role of the Ricci tensor, while the rank four tensor $A_{\alpha\beta\gamma\delta}$ would be analogous to the Weyl conformal tensor. While we could proceed with the analogy of curvature of spacetime to the ``curvature'' of the electromagnetic field, we will not pursue this topic further in the present paper. 

The fact that the two possible true invariants of the electromagnetic field are found as the real and imaginary parts of the complex scalar invariant is natural, since all inertial observers should agree upon their measured values.   
That the energy-momentum-stress tensor is found to be an irreducible component under SL(2,C) is encouraging and stress its importance in Maxwells theory. 

More surprising is the (pseudo)tensor $D_{\alpha\beta}$ found as the second irreducible rank two component. 
To the authors knowledge, this rank two tensor has not been written down explicitly in this form before\footnote{The authors learned that the new rank two tensor has been found independently by T.~D.~Carozzi by another method. Submitted to J. Math. Phys. (2005).}. We stress that this tensor is obtained on equal footing and occur as naturally as the energy-momentum stress tensor in our analysis. The components of $D_{\alpha\beta}$ represents observables, with a clear physical meaning of classical ``spin''.  
The new rank four tensor $C_{\alpha\beta\gamma\delta}$ is complex valued. Although the contracted imaginary part $\Upsilon_{\alpha\beta}$ is containing the imaginary part of the complex Poynting vector its physical interpretation is not yet clear and is leaved for a future study.

\section{Conclusions and Outlook}\label{seq:conclusions}
We have considered all possible bilinear forms of the electromagnetic Fourier transformed field for an arbitrary time-independent metric in Riemannian spacetime. The constructed spectral density tensor was decomposed into irreducible components by considering the spectral density spinor in the complex tangent space introduced under the group SL(2,C). The spectral density was found to be comprised of the following irreducible components: a complex scalar invariant corresponding to the two known invariants of the electromagnetic field, the rank two energy-momentum-stress tensor, another new rank two tensor expressing the spin (polarization) properties of the electromagnetic field, another rank two tensor, containing the imaginary part of the complex Poynting vector, and finally a new rank four tensor. This decomposition is expressed in Eqns~(\ref{eq:spinordecomptotal}) and (\ref{eq:irrtensorsymb}). While the first two tensors (scalar and stress-tensor) are well known important objects in Maxwells theory, the new rank two and rank four tensors are very little or not at all previously studied. Since they are irreducible components under the group transformations we propose that they also are conserved quantities of the electromagnetic field. 
Considering how they arise as naturally and inevitably as the ususal energy-momentum-stress tensor, they certainly deserve further study. We leave the detailed study of the discovered components and their conservation laws to a future investigation.  
\begin{acknowledgments}
The author thanks and acknowledge Dr. T.~D. Carozzi for useful discussions regarding the identfication of the tensors obtained, and for pointing out the need for 
a covariant formulation of polarization states and that spinors could be a possible way to accomplish it.
\end{acknowledgments}

\appendix

\section{Spinor representation of the group SL(2,C) in curved spacetime}\label{app:spinors}
In this appendix we review the theory of spinors and how they are applied
to general relativity and this article. In a Riemannian space one can 
to each point in space-time introduce a complex two-dimensional tangent space.
For every tensor in Riemannian space there is then a corresponding complex spinor in the tangent
spinor space.
For every tensor index labeled with greek letters running over 0, 1, 2, 3 
there are then two spinor indices labeled with Roman capital letters running over 0, 1.
Primed spinor indicies belong to the complex conjugate spinor space and 
and runs over $0^\prime, 1^\prime$.
Spinors originate from the representation theory of the group SL(2,C). We use a 
matrix representation in which a typical element g of the group SL(2,C) is represented
by the matrix
\begin{equation}
g=\left( 
\begin{array}{cc}
a & b\\
c & d
\end{array}
\right), \quad ad-bc=1
\end{equation}
with determinant unity.

The correspondence between tensors and spinors are established by the Infeld van der Waerden
symbols, a set of four 2x2 hermitian matrices $\sigma^\mu_{AB^\prime}(x^\nu)$, which are functions
of space-time. These objects transform as a tensor in greek indicies, and as a spinor in 
Roman indicies. They satisfy the relation
\begin{equation}\label{eq:infeld_metric_connection}
g_{\mu\nu}\sigma^\mu_{AB^\prime}\sigma^\nu_{CD^\prime}=\varepsilon_{AC}\varepsilon_{B^\prime D^\prime}
\end{equation}
relating the Infeld van der Waerden symbols to the metric tensor of Riemannian space and the Levi-Cevita
symbols, represented by
\begin{equation}
\varepsilon_{AC}=\varepsilon_{B^\prime D^\prime}=\varepsilon^{AC}=\varepsilon^{B^\prime D^\prime}=
\left( 
\begin{array}{rl}
0 & 1\\
-1 & 0
\end{array}
\right).
\end{equation}
The Levi-Cevita symbols play the spinor role analog to the metric tensors. The operation of 
raising and lowering 
spinor indicies are accomplished by 
\begin{equation}\label{eq:rasing_lowering_spinors}
\eta^A=\varepsilon^{AB}\eta_B,\quad \eta_A=\eta^B\varepsilon_{BA}
\end{equation}
and analogously for primed indicies
\begin{equation}\label{eq:rasing_lowering_spinorsprim}
\xi^{A'}=\varepsilon^{A'B'}\xi_{B'},\quad \xi_{A'}=\xi^{B'}\varepsilon_{B'A'}.
\end{equation}
One can easily check that the spinor of the metric tensor satisfies
\begin{equation}
g_{AB'CD'}=\varepsilon_{AC}\varepsilon_{B'D'}.
\end{equation}

The relation between a tensor $T_{\mu\nu}$ and the corrsponding spinor is given by
\begin{equation}\label{eq:spinor_tensor_connection}
T_{AB^\prime CD^\prime}=\sigma^\mu_{{AB^\prime}}\sigma^\nu_{{CD^\prime}}T_{\mu\nu}
\end{equation}
and analogously for tensors with more indicies.
If the tensor is real valued, the spinor equivalent is hermitian, e.g the real valued
vector $A_\mu$ have a equivalent hermitian spinor
\begin{equation}
A_{AB^\prime}=\overline{A_{BA^\prime}}=\overline A_{B^\prime A}.
\end{equation}
If the tensor is complex the hermicity condition of the spinor does not hold. This is the case
for most spinors encountered in this article.
Finally, given a spinor $T_{AB^\prime CD^\prime}$ its tensor equivalent $T_{\mu\nu}$
is given by 
\begin{equation}\label{eq:tensor_spinor_connection}
\sigma_\mu^{{AB^\prime}}\sigma_\nu^{{CD^\prime}}T_{AB^\prime CD^\prime}=T_{\mu\nu}.
\end{equation}

We also list the following useful spinor relations
\begin{equation}
\varepsilon^{AB}\varepsilon_{CA}=-\delta_C^{\phantom{C}B}
\end{equation}
\begin{equation}\label{eq:useful_relation1}
\xi_{AB}-\xi_{BA}=\xi_{C}^{\phantom C C}\varepsilon_{AB}=\varepsilon^{CD}\xi_{CD}\varepsilon_{AB}
\end{equation}
\begin{equation}\label{eq:useful_relation2}
\psi_{ABCD}-\psi_{BACD}=\psi_{F\phantom{F}CD}^{\phantom F F}\varepsilon_{AB}=\varepsilon^{FG}\psi_{FGCD}\varepsilon_{AB}
\end{equation}

\newpage 


\begin{thebibliography}{2}
\expandafter\ifx\csname natexlab\endcsname\relax\def\natexlab#1{#1}\fi
\expandafter\ifx\csname bibnamefont\endcsname\relax
  \def\bibnamefont#1{#1}\fi
\expandafter\ifx\csname bibfnamefont\endcsname\relax
  \def\bibfnamefont#1{#1}\fi
\expandafter\ifx\csname citenamefont\endcsname\relax
  \def\citenamefont#1{#1}\fi
\expandafter\ifx\csname url\endcsname\relax
  \def\url#1{\texttt{#1}}\fi
\expandafter\ifx\csname urlprefix\endcsname\relax\def\urlprefix{URL }\fi
\providecommand{\bibinfo}[2]{#2}
\providecommand{\eprint}[2][]{\url{#2}}

\bibitem[{\citenamefont{{Carmeli}}(1982)}]{carmeli1982a}
\bibinfo{author}{\bibfnamefont{M.}~\bibnamefont{{Carmeli}}},
  \emph{\bibinfo{title}{{Classical Fields: General Relativity and Gauge
  Theory}}} (\bibinfo{publisher}{{Wiley, New York}},
  \bibinfo{year}{1982}).

\bibitem[{\citenamefont{{Witten}}(1959)}]{witten1959a}
\bibinfo{author}{\bibfnamefont{L.}~\bibnamefont{{Witten}}},
  \bibinfo{journal}{Phys.~Rev.} \textbf{\bibinfo{volume}{113}},
  \bibinfo{pages}{357-362} (\bibinfo{year}{1959}).

\end{thebibliography}
\end{document}